\documentclass[preprint,aps,draft,showpacs]{revtex4}

\usepackage{graphicx}
\usepackage{dcolumn}
\usepackage{bm}

\begin{document}

\title{Dynamical Theory of Phase Transitions and
Topological Defect Formation in the Early Universe}

\author{Sang Pyo Kim}\email{sangkim@kunsan.ac.kr}

\affiliation{Department of Physics, Kunsan National University,
Kunsan 573-701, Korea}

\affiliation{Asia Pacific Center for Theoretical Physics, Pohang
790-784, Korea}

\date{\today}
\begin{abstract}
We review the current issues of nonequilibrium phase transitions,
in particular, in the early universe. Phase transitions cannot
maintain thermal equilibrium and become nonequilibrium when the
thermal relaxation time scale is greater than the dynamical time
scale. Nonequilibrium phase transitions would have happened in
certain evolution stages of the early universe because the rapid
expansion quenched matter fields. We apply the recently introduced
Liouville-von Neumann method, another canonical method, to
nonequilibrium phase transitions in the Minkowski spacetime and
find the scaling behavior of domain sizes. Topological defects are
thus determined by the dynamical processes of nonequilibrium phase
transitions. However, the expansion of the universe freezes domain
growth in the comoving frame and thus leads to a scale-invariant
domain size. We also discuss the physical implications of
nonequilibrium phase transitions in the early universe.
\end{abstract}

\maketitle

\section{Introduction}

A system undergoes a phase transition when its symmetry is broken
explicitly \cite{kibble,zurek,vilenkin}. The phase transition
proceeds either in equilibrium or in nonequilibrium (out of
equilibrium) depending on the ratio of the thermal relaxation time
to the dynamical time. When the thermal relaxation time scale is
shorter or longer than the dynamical time scale, the system
evolves in equilibrium or out of equilibrium. In particular, a
quenched system undergoes nonequilibrium phase transition when the
quench rate is faster than the relaxation rate. Matter fields are
believed to have undergone such nonequilibrium phase transitions
in the early universe as the universe expanded and thereby
temperature dropped rapidly. It is likely that such nonequilibrium
phase transitions of matter fields can also be realized and
observed in RHIC and LHC experiments in the near future.

The finite-temperature field theory has been the most popular
approach to equilibrium phase transitions \cite{dolan}. The
effective potential of quantum fluctuations around a classical
background provides a convenient tool to describe phase
transitions. However, quantum fluctuations of long wavelength
modes suffer from instability during the phase transition, become
unstable and grow exponentially. This is the origin of the
imaginary part of the effective action for phase transitions,
which gives the decay rate of the false vacuum \cite{weinberg}.
Thus the finite-temperature effective action does not properly
take into account the dynamical processes of phase transitions.

On the other hand, nonequilibrium phase transitions have been
frequently treated in the closed-time path integral defined in a
complex time plane by Schwinger \cite{schwinger} and Keldysh
\cite{keldysh}. Other methods are the Hartree-Fock or mean field
method \cite{ringwald,boyanovsky,rivers,cormier,destri}, the
$1/N$-expansion method \cite{cooper,destri2,baacke}, the
time-dependent variational principle \cite{balian}, and etc. In
this paper we use the recently developed Liouville-von Neumann
(LvN) method \cite{kim1,kim2,kim3,kim4,kim5}. The LvN method is a
canonical method trying to solve the functional Schr\"{o}dinger
equation \cite{freese,eboli}. The quantum LvN equation provides
invariant operators that solve exactly the functional
Schr\"{o}dinger equation \cite{lewis}. Compared with other
methods, this canonical method has advantages that it can be
applied directly to time-dependent systems and density operators
are readily found. The mode-decomposed Hamiltonian of quenched
models for phase transitions has this time-dependent feature.

We use quenched models of $\Phi^4$ theory with the sign changing
mass term for nonequilibrium phase transitions. These quenched
models have quench rates that determine explicitly time-dependent
phase transition processes, an instantaneous quench model
\cite{boyanovsky,rivers,kim2} and a finite quench model
\cite{kim2,bowick}. In both models the field, after symmetry
breaking phase transition, begins to roll from the false vacuum
toward the true vacuum. During this period of phase transition,
the system undergoes spinodal instability and its long wavelength
modes grow exponentially. Domains grow through spinodal
decomposition
\cite{boyanovsky,rivers,kim2,kim3,kim4,kim5,bowick,stephens}. Thus
domains and topological defects are dynamically determined by the
nonequilibrium field theory. Domain growth in the Minkowski
spacetime exhibits the Cahn-Allen scaling behavior of classical
theory. Further, unstable long wavelength modes lose quantum
coherence through interaction with stable short wavelength modes
and show classical correlation, thus achieve classicality.
Finally, we apply this nonequilibrium field theory to the early
expanding universe and study the domain growth and topological
defects. Domains remain frozen in the inflation era and have
scale-invariance. This affects the topological density in the
early universe.

The organization of this paper is as follows: In Sec. II
equilibrium phase transitions are briefly reviewed. In Sec. III we
discuss the Kibble-Zurek mechanism for the dynamics of equilibrium
processes. In Sec. IV some quench models are introduced for
nonequilibrium phase transitions. In Sec.V we introduce
nonequilibrium quantum field theory based on the functional
Schr\"{o}dinger equation. In Sec. VI the quantum LvN equation is
used to solve the functional Schr\"{o}dinger equation. In Sec. VII
we find the correlation functions during the phase transitions,
which determine the domain size. In Sec. VIII nonequilibrium phase
transitions are studied in an expanding universe. In Sec. IX  we
show how the classicality of classical correlation and decoherence
can be achieved for long wavelength modes.

\section{Equilibrium Phase Transitions}

To understand the symmetry breaking or restoration mechanism, we
consider a scalar field model with the potential
\begin{equation}
V (\phi) = \frac{m^2}{2} \phi^2 + \frac{\lambda}{4!} \phi^4.
\end{equation}
For a negative $m^2$ the symmetry of the system is broken.
However, quantum fluctuations around the true vacuum may restore
the symmetry when the temperature is high enough so that the
thermal energy can overcome the potential energy difference
between the true and false vacua. To find the thermal contribution
(correction) to the classical potential, we divide the field, $
\phi = \phi_c + \phi_f$, where $\phi_c$ is the classical
background field and $\phi_f$ denotes quantum fluctuations around
$\phi_c$. Then the effective potential is given by
\begin{eqnarray}
V_{e} (\phi_c) \equiv {\rm Tr} (\rho V) = \frac{1}{2} (m^2_R +
\frac{\lambda_R}{24} T^2) \phi_c^2 + \frac{\lambda_R}{4!}
\phi_c^4,
\end{eqnarray}
where $\rho$ is the density operator, and $m_R$ and $\lambda_R$
are renormalized coupling constants. For the broken symmetry $m^2
\rightarrow - m^2$, one has the potential
\begin{equation}
V_{e} (\phi_c)  = \frac{1}{2} (T^2 - T_c^2) \phi_c^2 +
\frac{\lambda_R}{4!} \phi_c^4.
\end{equation}
The system thus restores the broken symmetry when $T > T_c$
\cite{dolan}. In other words, the symmetry can be spontaneously
broken when the temperature drops below the critical temperature
$T_c$.

In the early stage of evolution after the Big Bang, the Universe
would have undergone a sequence of phase transitions as the
temperature dropped due to the expansion. A possible sequence of
phase transitions based on particle physics is the GUT  phase
transition at $T_c \approx 10^{14} - 10^{16}$ Gev, the EW
(electroweak) phase transition at $T_c \approx 10^{2}$ Gev, and
Color definement/Chiral phase transition at $T_c \approx 10^{2}$
Mev. Depending on the particle physics model, the system produces
different type of topological defects \cite{vilenkin}. The full
symmetry of the system is broken to a subgroup after a phase
transition. The structure of the vacuum manifold ${\cal M}$, the
homotopy group, determines the type of topological defects: domain
walls for $\pi_0 (\cal M) \neq 1$, strings for $\pi_1 (\cal M)
\neq 1$, and monopoles for $\pi_2 (\cal M) \neq 1$.

\section{Kibble-Zurek Mechanism}

Kibble used the principle of causality and the Ginzburg
temperature to find the correlation length of the same phase
(domain) in phase transitions \cite{kibble}. The Ginzburg
temperature is the one where the thermal energy is comparable to
the free energy of a correlated region
\begin{equation}
k_b T_c \approx \xi^3 (T_G) \Delta F(T_G),
\end{equation}
so that the field can overcome the potential barrier to jump to
other configurations. In this case topological defects lose
stability. This temperature restricts the size of correlation
length for stable defects. Topological defects are located along
the boundaries of correlated regions. Thus there is one monopole
per volume $\xi^3$ and the density of monopoles is given by
$\kappa / \xi^3$, and similarly one string per area $\xi^2$ and
the density by $\kappa/\xi^2$.

On the other hand, the Zurek mechanism incorporates the dynamics
of equilibrium processes \cite{zurek}. In the adiabatic cooling
(quenching) the equilibrium correlation length and the equilibrium
relaxation time increase, respectively, as
\begin{equation}
\xi = \xi_0 |\epsilon|^{-\nu}, \quad \tau = \tau_0 | \epsilon|^{-
\mu}
\end{equation}
where $\epsilon = (T_c - T)/T_c$. Here $\mu$ and $\nu$ are
model-dependent parameters. Also $\epsilon$ is related with the
quench rate as $\epsilon = t/\tau_Q$. As temperature approaches
the critical value $(T \rightarrow T_c$ or $t \rightarrow 0)$, the
equilibrium relaxation time becomes sufficiently longer and the
process critically slows down. However, the correlation length
increases indefinitely but the propagation of small fluctuations
is finite; so $\tau \propto \xi/v \rightarrow \infty$. Therefore,
there is a time $t^*$ when the correlation length freezes: $ |t^*|
= \tau (t^*)$. From the above equations the correlation length is
given by
\begin{equation} \xi(t^*) \approx \tau_Q^{\nu/(1 +
\mu)}.
\end{equation}
Thus the Kibble-Zurek mechanism determines the domain size
dynamically in terms of the quench rate.

\section{Nonequilibrium Phase Transitions}

There are many cases to which nonequilibrium phase transitions can
be applied but equilibrium phase transitions cannot be applied.
Matter fields in a rapidly expanding universe are such an example,
where the Tolman temperature drops as
\begin{equation}
T (t) = T_0 \frac{a_0}{a(t)},
\end{equation}
where $a$ is the scale factor of the universe. Another example is
the rapid quenching processes such as in the rapid cooling of
quark-gluon plasma in the Heavy Ion Collision and the liquid
helium ${\rm He}^3$ and ${\rm He}^4$, where domain walls and
vortices (strings) may be formed.

As a field-theoretical model for nonequilibrium phase transitions,
we consider the scalar field potential
\cite{boyanovsky,rivers,kim2,kim3,kim4,kim5,bowick}
\begin{equation}
V(\phi) = \frac{1}{2} m^2 (t) \phi^2 + \frac{\lambda}{4!} \phi^4,
\label{pot}
\end{equation}
where the mass $m^2$ changes signs from $ m^2 (- \infty) =
m^2_{i}$ to $ m^2 (+ \infty) = - m_{f}^2$. When $\tau_Q$ is the
time scale for the quench, one can distinguish the adiabatic
quench process, $\Delta m^2/ \tau_Q  \ll 1$, from the rapid quench
process, $\Delta m^2/ \tau_Q \gg 1$. For instance, an analytical
model may be considered \cite{kim2}
\begin{equation}
m^2 (t) = - m^2 \tanh \Bigl(\frac{t}{\tau_Q} \Bigr),
\end{equation}
 in which $m^2 \rightarrow + m^2$ for $t
\rightarrow - \infty$ and $m^2 \rightarrow - m^2$ for $t
\rightarrow + \infty$. In the limiting case $\tau_Q \rightarrow
0$, one has the instantaneous quench, $m^2 = + m^2$ for $t < 0$
and $m^2 = - m^2$ for $t > 0$. The field model can easily be
generalized to an expanding universe
\begin{equation}
H (t) = \int d^3 x \Bigl[\frac{\pi_{\phi}^2}{2 a^3(t)} + a^3(t)
\Bigl( \frac{(\nabla \phi)^2}{2 a^2(t)} + V(\phi) \Bigr) \Bigr],
\end{equation}
where $a$ is the scale factor of the FRW universe. The Minkowski
spacetime is the special case of $a = 1$. In the Hartree-Fock and
mean-field approximation, we divide $\phi = \phi_c + \phi_f$ to
obtain the equations of motion for the classical field
\begin{eqnarray}
\ddot{\phi}_c + 3 \frac{\dot{a}}{a} \dot{\phi}_c - \nabla^2 \phi_c
+ \Bigl(m^2 (t) + \frac{\lambda}{6} \phi_c^2 + \frac{\lambda}{2}
\langle \phi_f^2 \rangle \Bigr) \phi_c = 0, \label{cl eq}
\end{eqnarray}
and for quantum fluctuations
\begin{eqnarray}
\ddot{\phi}_f + 3 \frac{\dot{a}}{a} \dot{\phi}_f - \nabla^2 \phi_f
+ \Bigl( m^2 (t) + \frac{\lambda}{2} \phi_c^2 + \frac{\lambda}{4}
 \phi_f^2 \Bigr) \phi_f = 0. \label{qu eq}
\end{eqnarray}

\section{Nonequilibrium Quantum Field Theory}

The functional Schr\"{o}dinger-picture provides the real-time
evolution of (time-dependent) quantum systems \cite{freese,eboli}.
It is based on the functional Schr\"{o}dinger equation (in unit of
$\hbar = 1$)
\begin{equation}
i \frac{\partial}{\partial t} \Psi( \phi, t) = \hat{H} (\phi, - i
\frac{\delta}{\delta \phi}, t) \Psi(\phi, t), \label{sch eq}
\end{equation}
where $\phi$ represents a scalar or fermion field. It describes
the evolution of wave functionals $\Psi( \phi, t)$ from one
spacelike hypersurface $\Sigma_{t_0}$ to another $\Sigma_t$. The
set of all wave functionals constitutes a Hilbert space, which has
an inner product on each spacelike hypersurface $\Sigma_t$
\begin{equation}
\langle \Psi_1 \vert \Psi_2 \rangle = \int {\cal D} [\phi]
\Psi_1^* (\phi, t) \Psi_2 (\phi, t).
\end{equation}
The action of operators are defined as
\begin{equation}
\hat{{\cal O}} (\phi, \pi) \vert \Psi (\phi, \pi) \rangle
\rightarrow \hat{{\cal O}} ( \phi, - i \frac{\delta}{\delta \phi})
\Psi (\phi, t).
\end{equation}

Now the task is to solve Eq. (\ref{sch eq}) for the symmetry
breaking potential (\ref{pot}).  The evolution of the wave
functional may be found in terms of Green function (kernel or
propagator)
\begin{equation}
\Psi ({\bf x}, t) = \int G({\bf x}, t; {\bf x}_0, t_0) \Psi({\bf
x}_0, t_0) d{\bf x}_0 dt_0,
\end{equation}
where the wave functional implicitly depends on space through
$\phi$. Our stratagem is to separate the Hamiltonian into a
quadratic part, an exactly solvable one, and a perturbation part:
\begin{equation}
\hat{H} (t) = \hat{H}_0 (t) + \lambda \hat{H}_P (t).
\end{equation}
Here $\hat{H}_0$ includes not only the quadratic potential term
but also some contribution from non-linear terms a la the
Hartree-Fock or mean-field approximation. In this way the
approximation becomes nonperturbative including parts of quantum
corrections from the nonlinear terms. Then we may introduce a
Green function for $\hat{H}_0$ as
\begin{equation}
\Bigl(i \frac{\partial}{\partial t} - \hat{H}_0 ({\bf x}, t)
\Bigr) G_0 ({\bf x}, t; {\bf x}', t') = \delta ({\bf x} - {\bf
x}') \delta (t - t'), \label{gr fun}
\end{equation}
and write the wave functional in terms of $G_0$
\begin{equation}
\Psi ({\bf x}, t) = \Psi_0 ({\bf x}, t) + \lambda \int G_0 ({\bf
x}, t; {\bf x}', t') \hat{H}_P ({\bf x}', t') \Psi ({\bf x}', t')
d{\bf x}' dt', \label{gr wav1}
\end{equation}
where $\Psi_0$ is a wave functional for $\hat{H}_0$. The wave
functional $\Psi$ can be put recursively into the righthand side
of Eq. (\ref{gr wav1}) to result in
\begin{equation}
\Psi (1) = \Psi_0 (1) + \lambda \int G_0 (1,2) \hat{H}_P (2)
\Psi_0 (2)+ \lambda^2 \int \int G_0 (1,2) \hat{H}_P (2) G_0 (2,3)
\hat{H}_P (3) \Psi_0 (3) + \cdots, \label{gr wav2}
\end{equation}
where $(i)$ denotes $({\bf x}_i, t_i)$.

The general wave functional for $\hat{H}_0$ can take the Gaussian
wave functional
\begin{eqnarray}
\Psi_0 (\phi, t) &=& N \exp \Bigl[ - \int_{{\bf x}, {\bf y}} (\phi
({\bf x}) - \bar{\phi}({\bf x}, t)) \Bigl(\frac{1}{4 G ({\bf x},
{\bf y}, t)} - i \Sigma ({\bf x}, {\bf y}, t) \Bigr) \nonumber\\
&& \times (\phi ({\bf y}) - \bar{\phi}({\bf y}, t)) + i \int_{\bf
x} \bar{\pi} ({\bf x}, t) (\phi ({\bf x}) - \bar{\phi}({\bf x},
t)) \Bigr].
\end{eqnarray}
It has nonzero expectation values of the field and momentum
\begin{eqnarray}
\langle \Psi_0 \vert \hat{\phi} ({\bf x}) \vert \Psi_0 \rangle =
\bar{\phi} ( {\bf x}, t), \quad \langle \Psi_0 \vert \hat{\pi}
({\bf x}) \vert \Psi_0 \rangle = \bar{\pi} ( {\bf x}, t),
\end{eqnarray}
and also the two-point correlation functions
\begin{eqnarray}
\langle \Psi_0 \vert \hat{\phi} ({\bf x}) \hat{\phi} ( {\bf y})
\vert \Psi_0 \rangle &=& \bar{\phi}({\bf x}, t) \bar{\phi} ({\bf
y}, t) + G ({\bf x}, {\bf y}, t), \nonumber\\
\langle \Psi_0 \vert \hat{\pi} ({\bf x}) \hat{\pi} ( {\bf y})
\vert \Psi_0 \rangle &=& \bar{\pi}({\bf x}, t) \bar{\pi} ({\bf y},
t) + \Sigma ({\bf x}, {\bf y}, t).
\end{eqnarray}

\section{Canonical Method for Nonequilibrium Phase Transitions}

Recently, another canonical method, the so-called LvN or invariant
method, has been developed based on the quantum LvN equation
\cite{kim1,kim2,kim3,kim4,kim5}
\begin{equation}
i \frac{\partial}{\partial t} \hat{{\cal O}} (\phi, - i
\frac{\delta}{\delta \phi}, t)  + [ \hat{{\cal O}} (\phi, - i
\frac{\delta}{\delta \phi}, t), \hat{H} (\phi, - i
\frac{\delta}{\delta \phi}, t) ] = 0. \label{lvn eq}
\end{equation}
The idea of the LvN method for quantum mechanical systems first
exploited by Lewis and Riesenfeld \cite{lewis} is to solve Eq.
(\ref{lvn eq}) and find the solution to the Schr\"{o}dinger
equation as eigenstates of the operator in Eq. (\ref{lvn eq}). In
quantum field theory the wave functional to the Schr\"{o}dinger
equation is directly given by the wave functional of the operator
\begin{equation}
\hat{\cal O} ({\bf x}, t) \Psi_{\varphi} ({\bf x}, t) =
\varphi({\bf x}) \Psi_{\varphi} ({\bf x}, t).
\end{equation}
Note that the eigenvalue $\varphi ({\bf x})$ does not depend on
time, which is a consequence of Eq. (\ref{lvn eq}). In particular,
for the quadratic Hamiltonian the operator satisfying Eq.
(\ref{lvn eq}) can be obtained explicitly. This canonical method
has an advantage that quantum statistical information can be
naturally incorporated into the dynamics even for nonequilibrium
systems. Another merit is that it is a nonperturbative canonical
method and can further be applied to fermionic and gauge systems.

We now turn to the potential (\ref{pot}) for nonequilibrium phase
transition. As explained in Sec. V, we separate the Hamiltonian
density ${\cal H}$ into the quadratic part ${\cal H}_0$ and the
perturbation part ${\cal H}_P$:
\begin{eqnarray}
{\cal H}_0 (t) &=& \frac{1}{2 a^3} \pi^2 + \frac{1}{2a} (\nabla
\phi)^2 + \frac{a^3}{2} \Bigl(m^2 + \frac{\lambda}{2} \langle
\phi^2 \rangle
\Bigr) \phi^2, \nonumber\\
{\cal H}_P (t) &=& a^3 \Bigl(\frac{1}{4!} \phi^4 - \frac{1}{4}
\langle \phi^2 \rangle \phi^2 \Bigr).
\end{eqnarray}
It is convenient to decompose the field into Fourier modes and
then to redefine them as
\begin{equation}
\phi_{\bf k}^{(+)} (t) =  \frac{1}{\sqrt{2}} [ \phi_{\bf k} (t) +
\phi_{- {\bf k}} (t)], \quad \phi_{\bf k}^{(-)} (t) =
\frac{i}{\sqrt{2}} [ \phi_{\bf k} (t) - \phi_{- {\bf k}} (t)],
\end{equation}
where $\phi_{\bf k}^{(+)}$ and $\phi_{\bf k}^{(-)}$ are the
Fourier-cosine and sine modes, respectively.  For simplicity
reason a compact notation $\alpha, \beta, \cdots$, will be used
for $\{ {\bf k}, (\pm) \}$. Then the actual Hamiltonian of
symmetric state modes takes the form
\begin{equation}
H (t) = \sum_{\alpha} \Bigl[ \frac{1}{2a^3}\pi_{\alpha}^2 +
\frac{a^3}{2} \omega_{\alpha}^2 (t) \phi_{\alpha}^2 \Bigr] +
\frac{\lambda a^3}{4!} \Bigl[\sum_{\alpha} \phi_{\alpha}^4 + 3
\sum_{\alpha \neq \beta} \phi_{\alpha}^2 \phi_{\beta}^2 \Bigr],
\label{full ham}
\end{equation}
where
\begin{equation}
\omega_{\alpha}^2 (t) = m^2 (t) + \frac{{\bf k}^2}{a^2}.
\end{equation}
Thus the Hamiltonian consists of infinite number of coupled
anharmonic oscillators, all of which depend on time through
$m^2(t)$. The quadratic part of the Hamiltonian becomes
\begin{equation}
H_0 (t) = \sum_{\alpha} \frac{1}{2 a^3}\pi_{\alpha}^2 +
\frac{a^3}{2} \Omega_{\alpha}^2 (t) \phi_{\alpha}^2, \label{quad
ham}
\end{equation}
where
\begin{equation}
\Omega_{\alpha}^2 (t) = m^2 (t) + \frac{{\bf k}^2}{a^2} +
\frac{\lambda}{2} \sum_{\beta} \langle \phi_{\beta}^2 \rangle.
\end{equation}
Note that the Hamiltonian (\ref{quad ham}) simply consists of
decomposed oscillators.

We first find the Green function $G_0$ for $\hat{H}_0$ and then
obtain perturbatively the wave functional for the full
Hamiltonian. In fact, each mode of the quadratic part $\hat{H}_0$
can be solved exactly in terms the time-dependent creation and
annihilation operators satisfying the LvN equation. Then the wave
functional for $\hat{H}_0$ is the product of the wave function for
each mode. The time-dependent creation and annihilation operators
of the $\alpha$ mode are given by  \cite{kim1,kim2,kim3,kim4,kim5}
\begin{eqnarray}
\hat{a}_{\alpha}^{\dagger} (t)  = -i [ \varphi_{\alpha} (t)
\hat{\pi}_{\alpha} - a^3 \dot{\varphi}_{\alpha} (t)
\hat{\phi}_{\alpha}], \quad \hat{a}_{\alpha} (t)  = i [
\varphi_{\alpha}^* (t) \hat{\pi}_{\alpha} - a^3
\dot{\varphi}^*_{\alpha} (t) \hat{\phi}_{\alpha}]. \label{cr an}
\end{eqnarray}
We require these operators to satisfy the LvN equation with
$\hat{H}_{\alpha}$
\begin{eqnarray}
i \frac{\partial}{\partial t} \hat{a}^{\dagger}_{\alpha} (t) +
[\hat{a}^{\dagger}_{\alpha} (t), \hat{H}_{\alpha} (t)] = 0, \quad
i \frac{\partial}{\partial t} \hat{a}_{\alpha} (t) +
[\hat{a}_{\alpha} (t), \hat{H}_{\alpha} (t)] = 0.
\end{eqnarray}
Then the auxiliary field $\varphi_{\alpha}$ satisfies the
mean-field equation
\begin{equation}
\ddot{\varphi}_{\alpha} (t) + 3 \frac{\dot{a}(t)}{a(t)}
\dot{\varphi}_{\alpha} (t) + \Omega_{\alpha}^2 (t)
\varphi_{\alpha} (t) = 0. \label{aux eq}
\end{equation}
Indeed, these operators satisfy the usual commutation relations at
equal times
\begin{equation}
[ \hat{a}_{\alpha} (t), \hat{a}^{\dagger}_{\beta} (t) ] =
\delta_{\alpha \beta},
\end{equation}
when the Wronskian condition meets
\begin{equation}
a^3 (\dot{\varphi}^*_{\alpha} \varphi_{\alpha} -
\varphi^*_{\alpha} \dot{\varphi}_{\alpha}) = i. \label{wr con}
\end{equation}
In the oscillator representation of $\{
\hat{a}_{\alpha}^{\dagger}, \hat{a}_{\alpha} \}$, the quadratic
part takes the form
\begin{eqnarray}
\hat{H}_0 (t) &=& \frac{a^3}{2} \sum_{\alpha}
(\dot{\varphi}_{\alpha}^{2} + \Omega^2_{\alpha}
\varphi_{\alpha}^{2})\hat{a}^{2}_{\alpha} + 2
(\dot{\varphi}_{\alpha}^* \dot{\varphi}_{\alpha} +
\Omega_{\alpha}^2 \varphi_{\alpha}^* \varphi_{\alpha})
\hat{a}^{\dagger}_{\alpha} \hat{a}_{\alpha} +
(\dot{\varphi}^{*2}_{\alpha} + \Omega^2_{\alpha}
\varphi^{*2}_{\alpha})\hat{a}^{\dagger 2}_{\alpha},
\end{eqnarray}
and so does the perturbation part
\begin{eqnarray}
\hat{H}_P (t) &=& \frac{a^3}{4!}  \sum_{\alpha} \sum_{k = 0}^{4}
\Biggl[ {4 \choose k} \varphi_{\alpha}^{*(4-k)} \varphi_{\alpha}^k
\hat{a}^{\dagger (4-k) }_{\alpha} \hat{a}^k_{\alpha} + 3
\sum_{\alpha \neq \beta} \Bigl(\varphi_{\alpha}^2
\hat{a}^{2}_{\alpha} + 2 \varphi_{\alpha}^* \varphi_{\alpha}
\hat{a}^{\dagger}_{\alpha}\hat{a}_{\alpha} + \varphi_{\alpha}^{*2}
\hat{a}^{\dagger 2}_{\alpha}  \Bigr) \nonumber\\ && \times
\Bigl(\varphi_{\beta}^2 \hat{a}^{2}_{\beta} + 2 \varphi_{\beta}^*
\varphi_{\beta} \hat{a}^{\dagger}_{\beta}\hat{a}_{\beta} +
\varphi_{\beta}^{*2} \hat{a}^{\dagger 2}_{\beta}  \Bigr) \Biggr].
\label{per ham}
\end{eqnarray}

The essential point of the LvN method is that the number states of
$\hat{a}_{\alpha}^{\dagger}$ and $\hat{a}_{\alpha}$
\begin{equation}
\hat{N}_{\alpha} (t) \vert n_{\alpha}, t \rangle_0 =
\hat{a}^{\dagger}_{\alpha} (t) \hat{a}_{\alpha} (t) \vert
n_{\alpha}, t \rangle_0 = n_{\alpha} \vert n_{\alpha}, t
\rangle_0,
\end{equation}
are exact quantum states of the time-dependent Schr\"{o}dinger
equation. The quantum state of the field itself is then a product
of each mode state. For instance, the Gaussian vacuum state of the
field is given by
\begin{equation}
\vert 0, t \rangle_0 = \prod_{\alpha} \vert 0_{\alpha}, t
\rangle_0. \label{gauss vac}
\end{equation}
We now find the Green function for $\hat{H}_{\alpha}$
\begin{equation}
G_{0 \alpha} (\phi_{\alpha}, t; \phi_{\alpha}', t')  =
\sum_{n_{\alpha}} \langle \phi_{\alpha} \vert n_{\alpha}, t
\rangle_0 ~{}_0\langle n_{\alpha}, t' \vert \phi_{\alpha}'
\rangle,
\end{equation}
and the Green function for $\hat{H}_0$
\begin{equation}
G_0 ({\bf x}, t; {\bf x}', t' \rangle = \prod_{\alpha} G_{0
\alpha} (\phi_{\alpha}, t; \phi_{\alpha}', t'). \label{gr fn1}
\end{equation}
The wave functional for the full Hamiltonian (\ref{full ham}) can
be obtained at least perturbatively by substituting Eqs. (\ref{gr
fn1}) and (\ref{per ham}) into Eq. (\ref{gr wav2}).

\section{Domain Growth in Minkowski Spacetime}

We now study how the dynamical processes of nonequilibrium phase
transitions affect the domain growth and topological defect
density. The quench models in Sec. IV  describe such nonequilibrium
processes, which can be treated exactly for the free theory and
approximately for the self-interacting theory. Before the onset of
the phase transitions, the mass term dominates over the last term
from the quantum corrections. It is thus justified to use
approximately the free theory after the onset of phase transition
until it crosses the inflection point or spinodal line. This is
also true for the quench process lasting for an indefinitely long
period. We consider first the free theory in the Minkowski
spacetime and then discuss the effect of nonlinear terms.

The free theory for the quench models is provided by the potential
(\ref{pot}), where $\lambda = 0$ and $m^2(t)$ changes signs either
instantaneously or for a finite period. In the Minkowski spacetime
we can apply the formalism in Sec. VI simply by letting $a = 1$.
Before the phase transition $(m^2 = m_i^2)$, all the modes are
stable and oscillate around the true vacuum:
\begin{equation}
\varphi_{i 0 {\bf k}} (t) = \frac{1}{\sqrt{2 \omega_{i {\bf k}}}}
e^{- i \omega_{i {\bf k}}t}, \quad \omega_{i {\bf k}} = \sqrt{{\bf
k}^2 + m^2_i}.
\end{equation}
The two-point correlation function is the Green function at equal
times
\begin{equation}
G_0 ({\bf x}, {\bf x}', t) = \langle \hat{\phi} ({\bf x}, t)
\hat{\phi} ({\bf x}', t) \rangle_0 = G_0 ({\bf x}, t; {\bf x}', t)
\end{equation}
with respect to the Gaussian vacuum or thermal equilibrium. The
two-point thermal correlation function has the well-known form
\begin{eqnarray}
G_{i 0 T} ({\bf y}, {\bf x}, t) &=& \frac{1}{4 \pi^2} \frac{m_i
K_1 (m_i|{\bf x} - {\bf y}|)}{|{\bf x} - {\bf y}|} + \frac{m_i}{2
\pi^2 \sqrt{|{\bf x} - {\bf y}|^2 + m^2_i}} \nonumber\\
&& \times \sum_{n = 1}^{\infty} K_1[m_i\sqrt{|{\bf x} - {\bf y}|^2
+ (\beta n)^2}],
\end{eqnarray}
in terms of the modified Bessel function $K_1$. However, after the
phase transition $(m^2 = - m_f^2)$, the true vacuum becomes the
false vacuum.

First, in the instantaneous quench model, long wavelength modes
$(k < m_f)$ become unstable and exponentially grow as
\begin{equation}
\varphi_{f0{\bf k}} (t) = \frac{1}{\sqrt{2 \omega_{i{\bf k}}}}
\Bigl[ - i \frac{\omega_{i{\bf k}}}{\tilde{\omega}_{f{\bf k}}}
\sinh( \tilde{\omega}_{f{\bf k}}t) + \cosh( \tilde{\omega}_{f{\bf
k}}t) \Bigr], \quad \Bigl( \tilde{\omega}_{f {\bf k}} =
\sqrt{m^2_f - {\bf k}^2} \Bigr),
\end{equation}
whereas short wavelength modes $(k > m_f)$ are still stable and
have the solutions
\begin{equation}
\varphi_{f0{\bf k}} (t) = \frac{1}{\sqrt{2 \omega_{i{\bf k}}}}
\Bigl[ - i \frac{\omega_{i{\bf k}}}{\tilde{\omega}_{f{\bf k}}}
\sin( \tilde{\omega}_{f{\bf k}}t) + \cos( \tilde{\omega}_{f{\bf
k}}t) \Bigr], \quad \Bigl( \tilde{\omega}_{f {\bf k}} = \sqrt{{\bf
k}^2 - m^2_f} \Bigr).
\end{equation}
The two-point thermal correlation function  $(r = |{\bf x} - {\bf
x}'|)$ is dominated by the long wavelength modes and is given by
\begin{equation}
G_{f 0 T} (r,t) \simeq G_{0 T} (0, t) \frac{\sin\Bigl(
\sqrt{\frac{m_f}{2t}} r \Bigr)}{\sqrt{\frac{m_f}{2t}} r} \exp
\Bigl( - \frac{m_f r^2}{8 t} \Bigr). \label{two 1}
\end{equation}
From Eq. (\ref{two 1}) follows the famous Cahn-Allen scaling
relation for the domain size
\begin{equation}
\xi_D (t) = \sqrt{\frac{8t}{m_f}}. \label{ca sc}
\end{equation}
This implies that domains grow according to Eq. (\ref{ca sc})
until most of long wavelength modes cross the inflection point and
sample over the true vacuum.

Second, we consider a finite quench model with a finite quench
period. Such a model is mimicked by the mass term
\begin{equation}
m^2 (t) = m^2_1 - m^2_0 \tanh \Bigl(\frac{t}{\tau} \Bigr).
\label{f qc}
\end{equation}
The modes with $k < m_f = \sqrt{m_0^2 - m_1^2}$ become unstable
after the phase transition and exponentially grow, but those with
$k > m_f$ are still stable and oscillate around the false vacuum.
Long after the phase transition, the unstable long wavelength
modes have the asymptotic solution
\begin{equation}
\varphi_{f {\bf k}} = \frac{\mu_{\bf k}}{\sqrt{2} (m_f^2 - {\bf
k}^2)^{1/2}} e^{(m_f^2 - {\bf k}^2)^{1/2} t} + \frac{\nu_{\bf
k}}{\sqrt{2} (m_f^2 - {\bf k}^2)^{1/2}} e^{- (m_f^2 - {\bf
k}^2)^{1/2} t},
\end{equation}
whereas the stable short wavelength modes oscillate around the
false vacuum. Here $\mu_{\bf k}$ and $\nu_{\bf k}$ depend on the
quench process and satisfy the relation $|\mu_{\bf k}|^2 -
|\nu_{\bf k}|^2 = 1$. The correlation function is then
approximately given by
\begin{equation}
G_{f 0} ({\bf x}, t; {\bf x}', t) \simeq \int_0^{m_f} \frac{d^3
{\bf k}}{(2 \pi)^3} e^{i {\bf k} \cdot ({\bf x} - {\bf x}' )}
|\mu_{\bf k}|^2 \frac{e^{2 \sqrt{m_f^2 - {\bf k}^2} t}}{2 ({\bf
k}^2 - m_f^2)}. \label{2 fn}
\end{equation}
The coefficients $\mu_{\bf k}$ are obtained from the exact
solutions of Eq. (\ref{aux eq}) in terms of the hypergeometric
function \cite{kim2}. Using the exact solutions, we obtain the
two-point thermal correlation function during the the quench $( -
\tau < t < \tau)$
\begin{equation}
G_{m,T} (r,t) \simeq G_{m,T} (0, t) \frac{\sin \Bigl(
\frac{\sqrt{\tau t}}{m_0} r \Bigr)}{\frac{\sqrt{\tau t}}{m_0} r}
\exp \Bigl( - \frac{r^2}{8 \frac{\sqrt{\tau t}}{m_0}} \Bigr).
\end{equation}
Now the Cahn-Allen scaling relation reads
\begin{equation}
\xi_D (t) = 2 \Bigl(\frac{2 \tau t}{m_0^2} \Bigr)^{1/4}.
\end{equation}
After the quench $(t \gg \tau)$, the two-point thermal correlation
function is given by
\begin{equation}
G_{f_U,T} (r,t) \simeq G_{f_U,T} (0, t) \frac{\sin\Bigl(
\sqrt{\frac{m_f}{2\tilde{t}}} r
\Bigr)}{\sqrt{\frac{m_f}{2\tilde{t}}} r} \exp \Bigl( - \frac{m_f
r^2}{8 \tilde{t}} \Bigr),
\end{equation}
where
\begin{equation}
 \tilde{t} = t - \frac{\tau^3}{8}
[\zeta(3) - 1] (\omega^2_{i, {\bf k}} + \tilde{\omega}^2_{f, {\bf
k}}).
\end{equation}
Then the Cahn-Allen scaling relation for the domain size takes
another form
\begin{equation}
\xi_D (t) = \sqrt{\frac{8\tilde{t}}{m_f}}.
\end{equation}
The power of the scaling relation is the same as the instantaneous
quench, except for a time-lag proportional to the cube of the
quench duration $\tau$. Surprisingly, the correlation function has
the pole structure at
\begin{equation}
\tilde{\omega}_{f, {\bf k}} \tau = n ~~ (n = 1, 2, 3, \cdots).
\end{equation}
This implies larger domains for certain quench rates $\tau$
\cite{kim2}.

Finally, we discuss the effect of nonlinear term to the domain
growth. As all wave functionals $\Psi_0$ of $\hat{H}_0$ are
readily found, Eq. (\ref{gr wav2}) leads to the wave functional
beyond the Hartree approximation. Putting the perturbation terms
(\ref{per ham}) into Eq. (\ref{gr wav2}), we first find the wave
functional of the form
\begin{equation}
\Psi = \Psi_0^{(0)} + \sum_{n = 1}^{\infty} \lambda^n
\Psi^{(n)}_0, \label{beyond}
\end{equation}
and then calculate the Green function using the wave functional
(\ref{beyond}). An important result is that the correlation length
beyond the Hartree approximation has an additional factor
\begin{equation}
\xi (t) = \sqrt{2n+1} \xi_D (t).
\end{equation}
This factor from the $n$th order contribution is a consequence of
multiple scattering among different unstable modes \cite{kim5}.
The physical implication of the nonlinear effects is that the
correlation length increases by $(2n + 1)^{1/2}$, where $n$ is the
order of quantum contributions which depends the time for crossing
the spinodal line, i.e., the period for the field rolling from the
false vacuum into the true one.

\section{Topological Defect Density in an Expanding Universe}

As the unverse expands, the temperature drops and thus provides a
cooling process in a natural way. An adiabatically expanding
universe has the Tolman temperature
\begin{equation}
T (t) = T_c \frac{a_c}{a(t)},
\end{equation}
where $T_c$ is the critical temperature for the phase transition
and $a_c$ is the size at this moment. Then the mass term for the
second order phase transition may take the form
\begin{equation}
m^2(t) = T_c^2 \Bigl[\Bigl( \frac{a_c}{a(t)} \Bigr)^2 -1 \Bigr].
\label{ex mass}
\end{equation}
We now consider the free theory with the mass term (\ref{ex mass})
in the expanding universe. The Hamiltonian for the free field
takes the form
\begin{equation}
H_0 (t) = \sum_{\alpha} \frac{\pi_{\alpha}^2}{2 a^3(t)} +
\frac{a^3(t)}{2} \omega_{\alpha}^2 (t) \phi_{\alpha}^2,
\end{equation}
where
\begin{equation}
\omega_{\alpha}^2 (t) =  - T_c^2 + \frac{k^2 + T_c^2
a_c^2}{a^2(t)}.
\end{equation}

To find exact quantum states, we follow the formalism in Sec. VI.
The time-dependent creation and annihilation operators (\ref{cr
an}) are determined by the auxiliary variables now satisfying the
equation
\begin{equation}
\ddot{\varphi}_{\alpha} (t) + 3 \frac{\dot{a}(t)}{a(t)}
\dot{\varphi}_{\alpha} (t) + \omega^2_{\alpha} (t)
\varphi_{\alpha} (t) = 0. \label{v eq}
\end{equation}
Rewriting the auxiliary variables as $\varphi_{\alpha} = a^{-3/2}
v_{\alpha}$, Eq. (\ref{v eq}) can be written in a canonical form
\begin{equation}
\ddot{v}_{\alpha} (t) + \Bigl( - M^2 (t) + \frac{k^2 + T_c^2
a_c^2}{a^2(t)} \Bigr) v_{\alpha} (t) = 0,
\end{equation}
where
\begin{equation}
M^2 (t) = T_c^2 + \frac{3}{4} \Bigl(\frac{\dot{a}}{a} \Bigr)^2 +
\frac{3}{2} \frac{\ddot{a}}{a}.
\end{equation}
In particular, we are interested in the unstable long wavelength
modes with
\begin{equation} k < \sqrt{M^2 a^2 - T_c^2 a_c^2}.
\end{equation}
These long wavelength modes have both growing and damping
solutions of the form
\begin{equation}
v_{\alpha} = \frac{c_{1 \alpha}}{\sqrt{2 \Theta_{\alpha}}} e^{+
\int \Theta_{\alpha}} + \frac{c_{2 \alpha}}{\sqrt{2
\Theta_{\alpha}}} e^{- \int \Theta_{\alpha}}, \label{ad sol}
\end{equation}
where
\begin{equation}
\Theta_{\alpha}^2 (t) = M^2 (t) - \frac{k^2 + T_c^2 a_c^2}{a^2(t)}
- \frac{3}{4}
\Bigl(\frac{\dot{\Theta}_{\alpha}(t)}{\Theta_{\alpha}(t)} \Bigr)^2
+ \frac{1}{2}
\frac{\ddot{\Theta}_{\alpha}(t)}{\Theta_{\alpha}(t)}.
\end{equation}
Here $c_{1 \alpha}$ and $c_{2 \alpha}$ must satisfy $c_{1
\alpha}^* c_{2 \alpha} - c_{1 \alpha} c_{2 \alpha}^* = 1$ to
guarantee the Wronskian condition (\ref{wr con}). Assuming a slow
variation of $\Theta_{\alpha}$, we obtain the adiabatic (WKB)
solution with the exponent
\begin{equation}
\int \Theta_{\alpha} (t) = \int M (t) - \frac{k^2}{2} \int
\frac{1}{M(t) a^2 (t)} + \cdots.
\end{equation}
Then the adiabatic solutions lead to the correlation function
\begin{eqnarray}
G_{f 0} ({\bf x}, t; {\bf x}', t) &\simeq& \frac{1}{a^3 }\int
\frac{d^3 {\bf k}}{(2 \pi)^3} \Biggl[ e^{i {\bf k} \cdot ({\bf x}
- {\bf x}' )} \frac{|c_{1{\bf k}}|^2 }{2 \sqrt{M^2 - \frac{k^2 +
T_c^2
a_c^2}{a^2}}} \nonumber\\
&& \times  \exp \Bigl( 2 \int M - k^2 \int \frac{1}{M a^2} \Bigr)
\Biggr]. \label{ex gr}
\end{eqnarray}

Compared with the correlation function (\ref{2 fn}) in the
Minkowski spacetime, the correlation function (\ref{ex gr}) has an
overall decreasing factor $1/a^3$ from the expansion of the
universe. Further, after integrating over momenta, the Cahn-Allen
scaling relation for the domain size in the comoving frame is
given by
\begin{equation}
\xi_D (t) =  \Bigl( 8 \int_{t_0}^{t} \frac{1}{M a^2} \Bigr)^{1/2},
\end{equation}
where $t_0$ is the time for most of long wavelength modes to
undergo rolling motion from the false vacuum to the true vacuum.
For instance, during the inflation period with $a = e^{H_0 t}$,
the scaling relation becomes
\begin{equation}
\xi_D (t) = \Bigl[ \frac{4}{M_0 H_0} \Bigl( 1 - e^{-2 H_0 (t -
t_0)} \Bigr) \Bigr]^{1/2} e^{- H_0 t_0}, \label{ex dom}
\end{equation}
where
\begin{equation}
M_0^2 = T_c^2 + \frac{9}{4} H_0^2.
\end{equation}
The physical implication of Eq. (\ref{ex dom}) is that the domain
size in the comoving frame is frozen, $\xi_D (t) \approx \xi_D
(t')$ during the inflation period, in contrast with the Minkowski
spacetime where domains increase due to spinodal instability. Thus
there is no Cahn-Allen scaling relation and domains remain
scale-invariant in the comoving frame. However, the physical size
of domains increases in proportion to the scale factor of the
expanding universe
\begin{equation}
\xi_{PD} (t) = a (t) \xi_D.
\end{equation}
Of course, the higher order contributions beyond the Hartree
approximation increase both the comoving domain size and physical
domain size by additional factor $\sqrt{2n +1}$ for the $n$-th
order.

\section{Classicality of Quantum Phase Transitions}

We investigated phase transitions within the framework of quantum
field theory whereas phase transitions are studied in classical
theory in most literature. Though some of characteristic features
of phase transitions are well described and tested by classical
theory, quantum theory is more fundamental than classical theory.
So there should be some mechanism for the emergence of classical
features from quantum phase transitions. In more general context
than phase transitions, this is known as the quantum-to-classical
transition \cite{zeh}. In the previous section we have shown that
long wavelength modes play a key role in the dynamical process of
phase transitions and thus determine the growth of domains. In
classical theory long wavelength modes constitute just an order
parameter, which describes the dynamical process. For an order
parameter to achieve classicality, the quantum states of long
wavelength modes must lose quantum coherence and be classically
correlated along their classical trajectories
\cite{kim3,lombardo}.

To answer this question we consider a simple model motivated by
the potential (\ref{pot}). An exactly solvable model for a long
wavelength mode coupled to a short wavelength mode is provided by
the quadratic Hamiltonian \cite{kim3}
\begin{equation}
H (t) =  \frac{1}{2} \pi_1^2 + \frac{1}{2} {\omega}_1^2 (t)
\phi_1^2 + \frac{1}{2} \phi_2^2 + \frac{1}{2} {\omega}_2^2 (t)
\phi_2^2 + \lambda \phi_1 \phi_2. \label{co ham}
\end{equation}
Here $\phi_1$ and $\phi_2$ are the long and short wavelength
modes, respectively. The former becomes unstable but the latter
remains stable even after the phase transition. The time-dependent
Schr\"{o}dinger equation has a Gaussianal wave function of the
form
\begin{equation}
\Psi_0 (\phi_{1}, \phi_{2}, t) = N(t) \exp \Bigl[- \{A_{1} (t)
\phi_{1}^2 + \lambda B (t) \phi_{1} \phi_{2} + A_{2} (t)
\phi_{2}^2 \} \Bigr].
\end{equation}
Here $N$ is the normalization constant and the coefficients of the
exponent are given by
\begin{eqnarray}
A_{1}  (t) = -i\frac{\dot{u}_1^* (t)}{2 u_1^* (t)}, \quad A_{2}
(t) = - i \frac{\dot{u}_2^* (t)}{2 u_2^* (t)}, \quad B(t) = i
\frac{\int u_1^* (t) u_2^* (t)}{u_1^*(t) u_2^*(t)},
\end{eqnarray}
where
\begin{eqnarray}
\ddot{u}_1 (t) + \Bigl[\omega_{1}^2 (t) + \lambda^2 \Bigl(
\frac{\int u_1 (t) u_2(t)}{u_1(t) u_2(t)} \Bigr)^2
\Bigr] u_1 (t) = 0, \label{mx eq1}\\
\ddot{u}_2 (t) + \Bigl[ \omega_{2}^2 (t) + \lambda^2 \Bigl(
\frac{\int u_1 (t) u_2 (t)}{u_1 (t) u_2(t)} \Bigr)^2 \Bigr] u_2
(t) = 0. \label{mx eq2}
\end{eqnarray}

To model the instantaneous quench in Sec. VII, we assume $\omega_1^2
(t) = - \bar{\omega}_1^2$ and $\omega_2^2 (t) = \bar{\omega}^2_2$.
In the weak coupling limit $(\lambda \ll \bar{\omega}_1,
\bar{\omega}_2)$, we find the approximate solutions
\begin{equation}
u_1 (t) = c_1 e^{\bar{\Omega}_1 t}, \quad u_2 (t) = c_2 e^{ - i
\bar{\Omega}_2 t},
\end{equation}
where
\begin{eqnarray}
\bar{\Omega}_1 = \bar{\omega}_{1} - \frac{\lambda^2}{2
\bar{\omega}_{1}} \Bigl(\frac{\bar{\omega}_{1} + i
\bar{\omega}_{2}}{\bar{\omega}_{1}^2 + \bar{\omega}_{2}^2}
\Bigr)^2 - \frac{\lambda^4}{2 \bar{\omega}_{1}^2} \Bigl( i
\frac{1}{\bar{\omega}_{2}} + \frac{1}{4 \bar{\omega}_{1}} \Bigr)
\Bigl(\frac{\bar{\omega}_{1} + i
\bar{\omega}_{2}}{\bar{\omega}_{1}^2 + \bar{\omega}_{2}^2}
\Bigr)^4, \nonumber\\ \bar{\Omega}_2 = \bar{\omega}_{2} +
\frac{\lambda^2}{2 \bar{\omega}_{2}} \Bigl(\frac{\bar{\omega}_{1}
+ i \bar{\omega}_{2}}{\bar{\omega}_{1}^2 + \bar{\omega}_{2}^2}
\Bigr)^2 + \frac{\lambda^4}{2 \bar{\omega}_{2}^2} \Bigl( i
\frac{1}{\bar{\omega}_{1}} - \frac{1}{4 \bar{\omega}_{2}} \Bigr)
\Bigl(\frac{\bar{\omega}_{1} + i
\bar{\omega}_{2}}{\bar{\omega}_{1}^2 + \bar{\omega}_{2}^2}
\Bigr)^4.
\end{eqnarray}
Then the reduced density matrix for the long wavelength mode takes
the form
\begin{equation}
\rho_{\rm R} (\phi_1', \phi_1) = N^* N \sqrt{ \frac{\pi}{A_1^* +
A_1}} \exp \bigl[- \Gamma_{C} \phi_{1, C}^2 - \Gamma_{\Delta}
\phi_{1, \Delta}^2 - \Gamma_{M} \phi_{1, C} \phi_{1, \Delta}
\bigr],
\end{equation}
where $\phi_{1, C} = (\phi_1' + \phi_1)/2$ and $\phi_{1, \Delta} =
(\phi_1' - \phi_1)/2$. To order $\lambda^4$, the coefficients of
the exponent are approximated by
\begin{eqnarray}
\Gamma_{C} &=& \frac{5 \lambda^4 \bar{\omega}_{2}
(\bar{\omega}_{1} - \bar{\omega}_{2})}{2 \bar{\omega}_{1}
(\bar{\omega}_{1}^2 + \bar{\omega}_{2}^2)^4 }, \\ \Gamma_{\Delta}
&=& \frac{\lambda^2 }{\bar{\omega}_{2} (\bar{\omega}_{1}^2 +
\bar{\omega}_{2}^2)} \Bigl[1 + \frac{\lambda^2
\bar{\omega}_{2}}{(\bar{\omega}_{1}^2 + \bar{\omega}_{2}^2)^3}
\Bigl( \frac{\bar{\omega}_{1}^4}{2 \bar{\omega}_{2}^3} -
\frac{3\bar{\omega}_{1}^2}{\bar{\omega}_{2}} - \frac{5
\bar{\omega}_{2}^2 }{2 \bar{\omega}_{1}} + \bar{\omega}_{2} \Bigr)
\Bigr] ,
\\ \Gamma_{M} &=& - 2 i \bar{\omega}_{1} \Bigl[1 + \frac{\lambda^2
\bar{\omega}_{2}^2}{2 \bar{\omega}_{1}^2 (\bar{\omega}_{1}^2 +
\bar{\omega}_{2}^2)^2} + \frac{\lambda^4}{(\bar{\omega}_{1}^2 +
\bar{\omega}_{2}^2)^4} \Bigl(\frac{16}{5} - \frac{3
\bar{\omega}_{2}^2}{2 \bar{\omega}_{1}^2} -
\frac{\bar{\omega}_{2}^4}{4 \bar{\omega}_{1}^4}\Bigr) \Bigr].
\end{eqnarray}

Roughly speaking, a system is classically correlated when its wave
functions are peaked along classical trajectories or the contours
of the Wigner function is close to classical ones. And it
decoheres when each trajectory loses quantum coherence with its
neighbors. Quantum decoherence is realized when the diagonal term
$\phi_C$ of the density matrix dominates over the off-diagonal
term $\phi_{\Delta}$. A more precise measure of quantum
decoherence and classical correlation was introduced by Morikawa
\cite{morikawa}. According to his measure, quantum decoherence is
given by
\begin{equation}
\delta_{QD} = \frac{1}{2} \sqrt{\frac{\Gamma_C}{\Gamma_{\Delta}}}
= \frac{\lambda}{2}
\sqrt{\frac{5\bar{\omega}_{2}^2(\bar{\omega}_{1} -
\bar{\omega}_{2})}{2 \bar{\omega}_{1} (\bar{\omega}_{1}^2 +
\bar{\omega}_{2}^2)^2}},
\end{equation}
and classical correlation by
\begin{equation}
\delta_{CC} = \sqrt{\frac{\Gamma_c^2 \Gamma_{\Delta}^2}{\Gamma_M^*
\Gamma_M}} = \frac{5\lambda^6}{4} \frac{\vert \bar{\omega}_{1} -
\bar{\omega}_{2} \vert}{\bar{\omega}_{1}^2 (\bar{\omega}_{1}^2 +
\bar{\omega}_{2}^2)^5}.
\end{equation}
The system loses quantum coherence (decoheres) when $\delta_{QD}
\ll 1$ and recovers classical correlation when $\delta_{CC} \ll
1$.

We therefore find that the unstable long wavelength mode becomes
classical, gaining both quantum decoherence and classical
correlation due to the coupling to a stable short wavelength mode.
In field models, the infinitely large number of short wavelength
modes provides an environment to unstable long wavelength modes
and very quickly make them classical. Quite recently, field models
have been studied for the quadratic, Yukawa, and linear couplings
to the environment, where quantum decoherence of long wavelength
modes is observed and the decoherence time is found to be very
short compared with other time scales \cite{lombardo}. In this way
long wavelength modes become classical immediately after phase
transitions, from which appears a classical order parameter
\cite{kim3}.

\section{Conclusion}

We critically reviewed the nonequilibrium quantum phase
transitions by employing the canonical method called LvN
(Liouville-von Neumann) method. The nonequilibrium phase
transitions are characterized by the quench time scale shorter
than the thermal relaxation time scale, so that they cannot have
time to relax to thermal equilibrium and evolve out of
equilibrium. As such nonequilibrium phase transitions we have
considered the quenched second order phase transitions modelled by
the $\phi^4$-theory with an explicitly time-dependent mass that
changes sign during the quench process. This model is a quantum
model for the classical Ginzburg-Landau theory for spinodal
decomposition.

We used the LvN method to study such time-dependent quantum field
theory. The wave functionals of the Schr\"{o}dinger equation carry
all the quantum information of the system evolving from initial
data such as thermal equilibrium or Gaussian vacuum to final
nonequilibrium states. The essential idea of the LvN method is
first to solve the LvN equation in terms of whose operators exact
wave functionals are found for the time-dependent Schr\"{o}dinger
equation. In this LvN method it is easy to incorporate the thermal
equilibrium because density operators are ready to be found. In
particular, the exact quantum states and the Green function can be
found explicitly for a quadratic potential. So our stratagem is to
separate the Hamiltonian into a quadratic part of the Hartree-Fock
type and a perturbation. In the momentum space and the oscillator
representation, the perturbation part of our model consists only
of quartic terms of the creation and/ or annihilation operators.
By using the Green function for the quadratic part and solving
perturbatively the perturbation part, we are able to find  higher
order contributions to the wave functional. This method provides
to go beyond the Hartree-Fock approximation.

As a consequence of the direct application of the LvN approach to
the quenched second order phase transitions, we obtained some
interesting results such as growth of domain sizes, topological
density and decoherence of order parameter. In the instantaneous
quench model, the domain sizes scale according to the Cahn-Allen
scaling relation which has been observed in classical phase
transitions. In the finite quench model, the domain sizes have a
different scaling behavior during the quench. Surprisingly, the
correlation function after the quench exhibits resonance for
certain quench rates. Further, the nonlinear effect leads to a
multiple-scaling relation beyond the Hartree-Fock approximation.
Another interesting phenomenon is the classicality of phase
transitions due to instability of long wavelength modes. The long
wavelength modes become unstable, exponentially grow due to
spinodal instability and thus exhibit classical correlation. The
long wavelength modes therefore achieve classicality through the
coupling to a large number of stable short wavelength modes
(environment). It is found that the topological defect density can
be reduced by factors $(2n+1)^{3/2}$, which is a consequence of
multiple scattering at the higher order beyond the Hartree-Fock
approximation.

Quantum fields in the early unverse provide a natural framework
for nonequilibrium phase transitions as the adiabatically
expanding universe has the Tolman temperature dropping inversely
proportion to the scale factor of the universe. It is widely
believed that the universe would undergo a sequence of phase
transitions. However, the phase transitions may not be properly
treated in finite temperature field theory due to the adiabatic
expansion. We apply the LvN method to the quenched second order
phase transition in an expanding FRW universe. The damping effect
of field due to the expansion of the universe dramatically changes
the domain growth. In fact, domains in the comoving frame remain
frozen and show scale-invariant behavior in the inflation era.
This result strongly contrasts with the Cahn-Allen scaling
relation in the Minkowski spacetime, according to which domains
grow as a power of time due to spinodal instability. Thus domains
growing through dynamical processes suppress formation of
topological defects since topological defects form on the
boundaries of domains. However, the physical sizes of domains
increase in proportion to the scale factor of the universe. This
implies that topological defects may not be suppressed through
dynamical processes of nonequilibrium phase transitions but
through the inflation of the universe.

\acknowledgments 

The author would like to thank F. C. Khanna, C.
H. Lee and S. Sengupta for collaboration and useful discussions.
This work was supported by Korea Research Foundation under Grant
No. KRF-2002-041-C00053 and KRF-2003-041-C20053.




\begin{references}

\bibitem{kibble} T. W. B. Kibble, J. Phys. A {\bf 9},
1387 (1976).

\bibitem{zurek} W. H. Zurek, Nature {\bf 317}, 505 (1985).

\bibitem{vilenkin} A. Vilenkin and E. P. S. Shellard, {\it Cosmic
Strings and Other Topological Defects} (Cambridge Univ. Press,
Cambridge, 1994).

\bibitem{dolan} L. Dolan  and R. Jackiw, Phys. Rev. D
{\bf 9}, 3320 (1974).

\bibitem{weinberg} E. J. Weinberg  and A Wu, Phys. Rev. D {\bf
36}, 2474 (1987).

\bibitem{schwinger} J. Schwinger, J. Math. Phys. {\bf 2}, 407 (1961).

\bibitem{keldysh} L. V. Keldysh, Sov. Phys. JETP {\bf 20}, 1018 (1965).

\bibitem{ringwald} A. Ringwald, Phys. Rev. D {\bf 36}, (1987)
 2598.

\bibitem{boyanovsky} D.  Boyanovsky,   D.-S. Lee, and A. Singh,
Phys. Rev. D {\bf 48}, 800 (1993).

\bibitem{rivers} R. J.  Rivers, Int. J. Theor. Phys. {\bf 39}, 
1623 (2000).

\bibitem{cormier} D. Cormier and R. Holman, Phys. Rev. D
{\bf 62}, 023520 (2000).

\bibitem{destri} C. Destri and E. Manfredini, Phys. Rev. D
{\bf 62}, 025008 (2000).

\bibitem{cooper} F. Cooper, S. Habib, Y. Kluger, and E. Mottola,
Phys. Rev. D {\bf 55}, 6471 (1997).

\bibitem{destri2} C. Destri and E. Manfredini, Phys. Rev. D 
{\bf 62},  025007 (2000).

\bibitem{baacke} J. Baacke and K. Heitmann, Phys. Rev. D
{\bf 62},  105022 (2000).

\bibitem{balian} R.  Balian and M. V\'{e}n\'{e}roni, Ann.
Phys. (N.Y.) {\bf 164}, 334 (1985).


\bibitem{kim1} K. H. Cho, J.-Y. Ji, S. P. Kim, C. H. Lee,
and J. Y. Ryu, Phys. Rev. D {\bf 56}, 4916 (1997).

\bibitem{kim2} S. P. Kim and C. H.  Lee, Phys. Rev. D {\bf
62}, 125020 (2000).

\bibitem{kim3} S. P. Kim and C. H. Lee, Phys. Rev. D
{\bf 65}, 045013 (2002).

\bibitem{kim4}  S. P. Kim, S. Sengupta, and F. C. Khanna,
Phys. Rev. D {\bf 64}, 105026 (2001).

\bibitem{kim5} S. Sengupta, F. C. Khanna, and S. P. Kim, 
Phys. Rev. D {\bf 68}, 105014 (2003).

\bibitem{freese} K. Freese, C. T. Hill, and M. Mueller,
Nucl Phys. B {\bf 255}, 693 (1985).

\bibitem{eboli} O. \'{E}boli, R. Jackiw, and S.-Y. Pi,
Phys. Rev. D {\bf 37}, 3557  (1988).

\bibitem{lewis} H. R. Lewis, Jr. and W. B. Riesenfeld, J.
Math. Phys. {\bf 10},  1458 (1969).

\bibitem{bowick}  M. Bowick and A. Momen, Phys. Rev. D {\bf 58},
 085014 (1998).

\bibitem{stephens} G. J. Stephens, E. A. Calzetta, B. L. Hu, and
S. A. Ramsey, Phys. Rev. D {\bf 59},  045009 (1999).

\bibitem{zeh} D. Giulini, E. Joos, C. Kiefer, J. Kupsch,
I.-O. Stamatescu, and H. D. Zeh, {\it Decoherence and the
Appearance of a Classical World in Quantum Theory} (Springer,
Berlin, 1996).

\bibitem{lombardo} F. C. Lombardo,  F. D. Mazzitelli, and
D. Moteoliva, Phys. Rev. D {\bf 62}, 045016 (2000); F. C. Lombardo, 
F. D. Mazzitelli, and R. J. Rivers, Phys. Lett. B
{\bf 523},  317 (2001); F. C. Lombardo, F. D. Mazzitelli, 
and R. J. Rivers, Nucl. Phys. B {\bf 672}, 462 (2003).

\bibitem{morikawa} M. Morikawa, Phys. Rev. D {\bf 42}, 2929 (1990).


\end{references}
\end{document}